\documentclass{elsart}

\usepackage{graphicx}

\usepackage{amssymb}

\begin{document}

\begin{frontmatter}



\title{Two particle azimuthal correlations at high transverse momentum in Pb-Au at 158~AGeV/c}


\author{Mateusz P\l osko\'n for the CERES Collaboration}

\address{Institut f\"ur Kernphysik, J.W.Goethe Universit\"at, Frankfurt am Main, Germany}

\begin{abstract}
The analysis of two-particle azimuthal angular correlations at high transverse momentum from Pb+Au collisions at $158$~AGeV/c at SPS reveals substantial modifications of the away-side peak as compared to the distributions from p+p reactions. The data recorded with the CERES Time-Projection Chamber, which provides excellent tracking efficiency, suggest that the observed modification of the back-to-back structure implies significant re-interactions of the scattered partons within the medium. These findings at top SPS energy show features qualitatively similar to the results obtained at RHIC from Au+Au collisions at $\sqrt{s} = 200$~GeV/c and Cu+Cu collisions at $\sqrt{s} = 62$~GeV/c \cite{Adams:2005ph,PHENIX,McCumber:2005cv,Ulery:2005qm}. We present the centrality and charge dependent conditional di-jet yields for similar transverse  momentum windows as investigated by the RHIC experiments \cite{Adams:2005ph,PHENIX,McCumber:2005cv,Ulery:2005qm}.
\end{abstract}

\end{frontmatter}

\section{Motivation}
The strong supression of hadron yields at high transverse momentum in central heavy-ion collisions at RHIC, compared to yields in p+p scaled by the number of binary nucleon-nucleon collisions has been regarded as a manifestation of the extremely dense color charged medium created in the collision \cite{STAR,PHENIXdense}. Such a dense medium is expected to strongly enhance the energy loss of a hard-scattered parton and modify its fragmentation. A well suited tool to study jet production and to address their in-medium modification are azimuthal angular correlations with respect to a leading hadron. Already the pioneering studies at SPS have revealed broadening of the away-side in central Pb-Au collisions at $158$~AGeV/c \cite{JANA} and the recent studies at RHIC show further modifications to the correlation pattern \cite{Adams:2005ph,PHENIX,McCumber:2005cv,Ulery:2005qm}.
\section{Experiment and Data Analysis}
\begin{figure}[htb]
\begin{center}
\includegraphics[scale=0.65]{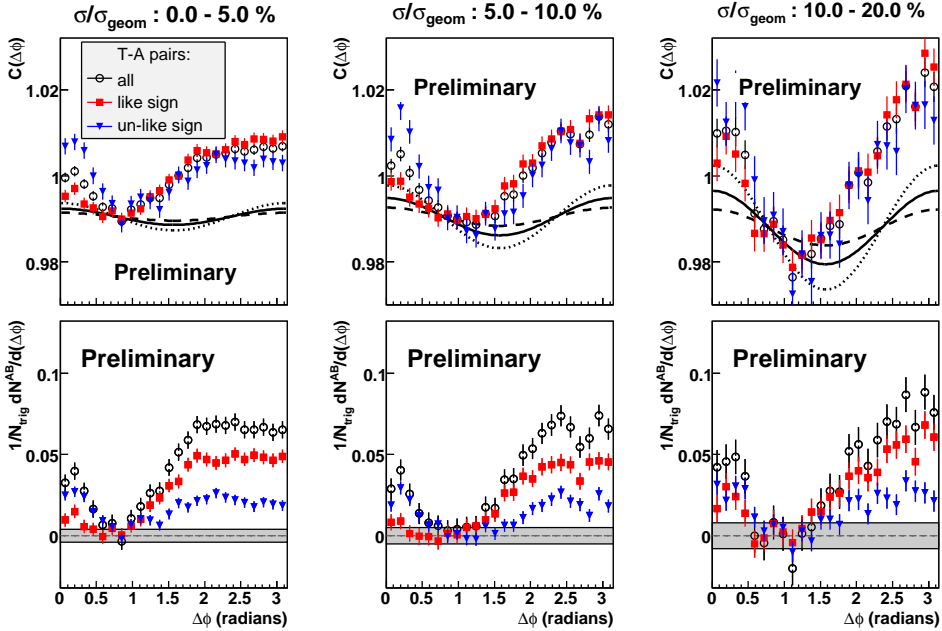}
\caption{\underline{\it Upper row:} Correlation functions of the like-sign and unlike-sign combinations of the trigger and associated particles for the three centrality bins. The open symbols show the correlation function w/o the charge selections. Solid line indicates the 2-particle flow contribution. The dashed and dotted lines represent the flow contributions with ($v^A_2$) and ($v^B_2$) varied according to their statistical uncertainties. \underline{\it Lower row:} Conditional yields of the jet-pair distributions for the three centrality bins, normalized to the number of triggers. The full, horizontal bands represent the systematic uncertainty resulting from the flow constant $b_0$ estimation by the ZYAM method.}
\label{fig1}
\end{center}
\end{figure}

The present study is based on the analysis of 30 million Pb-Au events at $158$~AGeV/c
recorded with the CERES spectrometer at SPS in the year 2000. The momenta of the charged
particles are determined by the curvatures of their tracks in the radial Time Projection Chamber
(TPC). The achieved momentum resolution is $\Delta p / p = ((2\%)^2 + (1\% p ($~GeV/c~$)^2)^{1/2}$ \cite{MISKO}. 
The detector acceptance is $2\pi$ in azimuth and $2.1$ to $2.7$ in pseudo-rapidity. 
The centrality and the number of participating nucleons have
been estimated with the nuclear overlap model calculation \cite{OVERLAP} using the charged particle multiplicities in the Silicon Drift Detectors (SDD) and the TPC. The data set has been subdivided into three centrality samples according to the $0-5\%$, $5-10\%$ and $10-20\%$ most central events, respectively.
\\
For the analysis of leading hadron correlations, we follow a scheme presented by the PHENIX collaboration \cite{PHENIX}. We select trigger particles in the range
$2.5$~GeV/c~$< p_t^{trigger} < 4.0$~GeV/c and calculate the azimuthal angular difference
$\Delta \phi^{same} $ with respect to associated particles with $1.0$~GeV/c~$< p_t^{assoc.} < 2.5$~GeV/c in the same event. 
The normalized yield $Y(\Delta \phi^{same})$ has been divided by the normalized
yield $Y(\Delta \phi^{mixed})$ obtained from mixed event pairs to form a correlation
function: $C(\Delta \phi) = \frac{Y_{same}(\Delta \phi)}{Y_{mixed}(\Delta \phi)} \cdot \frac{\int Y^{mixed}}{\int Y^{same}}$.  Due to the limited two-track resolution pairs of tracks with $\Delta \theta < 10$~mrad have been rejected in the same- and mixed event sample. The resulting correlation functions are shown in Fig. \ref{fig1}. The non-zero two-particle flow contribution ($1 + 2 < v_2^A v_2^B > \cos(2 \Delta \phi)$) has been approximated by ($ 1 + 2 < v_2^A > < v_2^B > \cos(2 \Delta \phi) $). The flow values  ($v_2^A$) for the trigger and ($v_2^B$) for associated particles
 have been obtained from the reaction plane method \cite{JANA,JOVAN}. The resulting flow
contributions are shown as solid lines in Fig. \ref{fig1}. In order to extract the (di-)jet signal,
the correlation function is decomposed into two contributions: one proportional to
the distribution of the background pairs (containing flow), and the second $J( \Delta \phi)$
representing the (di-)jet pairs:
$C(\Delta \phi) = b_{0} \cdot (1 + 2 \cdot < v^{A}_{2} > < v^{B}_{2} > \cos(2\Delta \phi)) + J(\Delta \phi)$.
The parameter $b_0$ is chosen to match the Zero Yield At Minimum
(ZYAM) condition \cite{PHENIX}. After the flow contribution is
subtracted from the correlation function, the fully corrected (di-)jet pair
distribution $dN^{AB}_{(Di-)Jet} / d\Delta \phi$  can be constructed. 
The conditional yield distribution of jet-associated particles per trigger is given by:
\begin{equation}\label{condyield}
\frac{1}{N_{trig}}\frac{dN^{AB}}{d\Delta \phi} = \frac{J(\Delta \phi)}{\int(C(\Delta \phi ')d(\Delta \phi '))} \frac{N^{AB}}{N^{A}},
\end{equation}
where $N^A$ is the number of triggers and $N^{AB}$ the total number of $AB$ pairs in
the event sample. The conditional yields have been corrected for the single hadron
efficiency estimated to be $85\%$ at laboratory momenta greater than $1.0$~GeV/c. This
correction leads to a $1.5\%$ systematic uncertainty on the absolute normalization.
Furthermore, the data sample has been divided into like-sign and unlike-sign pairs with respect to e.m. charge of the trigger and associated particle. Corresponding correlation functions are shown in the upper panel of the Fig. \ref{fig1}.
\section{Results and Summary}
The yields of the (di-)jet associated particles in the lower panel of the Fig. \ref{fig1} show significant modification of the away side as compared to the expected shapes from p+p collisions. Additionally, the away side structure in the central collisions develops a pronounced plateau, as compared to the less central sample, possibly indicating a local minimum at $\Delta\phi = \pi$ in the case of unlike-sign pairs. The substantial modification of the back-to-back structure is possibly caused by large re-interaction of the scattered partons in the medium. These findings, qualitatively similar to the observations at RHIC \cite{Adams:2005ph,PHENIX,McCumber:2005cv,Ulery:2005qm}, suggest that conceivable mechanisms such as conical flow emerging from supersonic partons in a thermalized colored medium \cite{STOECKER,SHURYAK} may also be at work at SPS. \\
Furthermore, the strength of the near side correlation for the unlike-sign pairs is larger than for the like-sing pairs. This may be due to charge ordering in the fragmentation process. On the other hand, the observed di-jet yield on the away-side is largest for the like-sign pairs. This behaviour is also observed in p+p events generated with PYTHIA at $\sqrt{s}=17$~GeV/c, however, the away-side yield difference between the like-sign and unlike-sign correlations vanishes at higher energies.
\\
The correlations presented here are under further investigations, also by means of the three-particle correlations.
 


\end{document}